\documentclass[aps,prd,preprint,showpacs,showkeys]{revtex4}
\usepackage{bm}
\usepackage[mathscr]{euscript}
\usepackage{braket}
\usepackage{amsmath}
\begin{document}

\title{Cosmological Birefringence and the Geometric Phase of Photons}

\author{Mansoureh Hoseini and Mohammad Mehrafarin}
\email{mehrafar@aut.ac.ir}
\affiliation{Physics Department, Amirkabir University of Technology, Tehran
15914, Iran}

\begin{abstract}
Regarding axion electrodynamics in the background flat FRW universe, we show that cosmological birefringence arises from an adiabatic noncyclic geometric phase that appears in the quantum state of photons because of their interaction with the axion field. We also show that the axion electrodynamics is equivalent to standard electrodynamics in time-dependent bi-isotropic magnetoelectric Tellegen media, which serves as an analogue system that can simulate cosmological birefringence.
\end{abstract}


\maketitle

\section{Introduction}
Dark matter, which neither emits nor scatters electromagnetic (EM) radiation, is considered to be one of the main cosmic  constituents of the universe, comprising some 23\% of its energy. The components of dark matter are  unknown, but there are several plausible candidates. One interesting class is furnished by massive pseudo-Goldstone bosons (axions and axion-like particles) \cite{Kolb1990,Marsh2016,Leon2017}; pseudo-scalar particles first proposed in order to resolve the strong CP problem in quantum chromodynamics \cite{Peccei1977}. A distinguishing feature of the axion field is that it couples to EM field, notably the cosmic microwave background (CMB) radiation. As a subject of growing interest, axion-electrodynamics, i.e., the standard electrodynamics modified by photon-axion interaction, provides a theoretical framework for possible violation of parity and Lorentz invariance  \cite{Carroll1990}. 

CMB radiation furnishes the best testing ground for axion-electrodynamics. Its polarization arises from the asymmetric Thomson scattering in the epoch of recombination, yielding the longest available look-back time as compared to any other cosmological source. Current measures and constraints on the polarization pattern of CMB anisotropies produce an upper limit on the linear polarization rotation angle of the order of few degrees (see, e.g., \cite{Wu2009,Chiang2010,Jarosik2011}). The EM-axion field interaction affects the polarization of EM waves propagating over cosmological distances \cite{Carroll1990,Carroll1991,Harari1992,Finelli2009}. In this connection,  Carroll et al. \cite{Carroll1990,Carroll1991} were first to show that different polarizations of  EM wave propagate in the same direction with different phases, causing a rotation of the polarization plane (cosmological birefringence). Comparison of the polarization rotation (which depends on the axion field features) with the CMB polarization observations, can give experimental confirmation of the existence (or otherwise) of axions and their physical properties.

Because of the dependence on polarization of  optical and transverse acoustical wave propagation in inhomogeneous media, there occurs similar polarization rotation (the Rytov rotation), which is established to be a manifestation of the Berry phase of the quanta (photons/phonons) of the quantized wave fields  \cite{Torabi2012,Torabi2008,Mehrafarin2009}. In this work, we show that the cosmological birefringence likewise arises from an adiabatic noncyclic geometric phase that appears in the quntum state of the photons because of their interaction with the slowly varying pseudo-scalar field. This phase, which is gauge and reparametrization invariant, is associated with the image of the path traversed by the state vector of photons in the projective Hilbert space. A noncyclic geometric phase also has been shown to characterize the mixing phenomenon of photons and axions in the presence of strong magnetic fields \cite{Capolupo}.

Berry, in his seminal work \cite{Berry1984}, showed that in a cyclic adiabatic quantum evolution, the wave function ends up in the original state up to a phase factor that is the sum of a dynamical phase and an additional geometric phase, known as the Berry phase. This phase has been generalized to the case of nonadiabatic \cite{Aharonov1987}, nonunitary, and noncyclic evolutions \cite{Samuel1988}. Also, using a quantum kinematic approach and the Pancharatnam criterion \cite{Pancharatnam1956}, Mukunda and Simon showed that a gauge and reparametrization invariant geometric phase can be naturally associated with any smooth open curve traversed by unit vectors in Hilbert space \cite{Mukunda1993}. Since its discovery, examples of geometric phase have appeared in many different areas of physics \cite{Chruscinski2004,Baggio2017}. It is now well known that geometric phases arise in systems with time-dependent Hamiltonians and non-stationary quantum states \cite{Zeng1995}.

Generalized time-dependent harmonic oscillator (TDHO) is a ubiquitous time-dependent quantum systems, which is apt for the application of the dynamical invariant method introduced by Lewis and Riesenfeld (LR) \cite{Lewis1969}.
In the present work, we show that axion electrodynamics in flat Friedmann-Robertson-Walker (FRW) universe, when quantized, can be described by a collection of  generalized TDHO Hamiltonians for the photons. We use the LR invariants to find the phase of the quantum state of photons interacting with the axion field.  By analyzing the phase factor, we then establish that cosmological birefringence arises from the geometric part of the phase.

 Finally, we present the axion electrodynamics from another perspective. We propose a similar system where the relevant features of photon-axion mixing in expanding spacetime can be reproduced analogously. This approach, known as analogue gravity \cite{Barcelo2011}, is the study of phenomena in curved spacetimes by means of analogue systems whose kinematics are governed by the same mathematical formalisms. With analogue models, it is possible to simulate predicted phenomena that have not yet been experimentally observed. They, therefore, provide a powerful tool for testing cosmological models. The analogue model that we propose to use for simulating cosmological birefringence is light traveling in time-dependent Tellegen medium \cite {Tellegen1948}. We show that the propagation of light in time-dependent bi-isotropic magnetoelectric  media shares the same behavior as in axion electrodynamics in flat FRW universe. The results obtained from this simulation can, thus, provide a  substitution for direct experimental observation.

The organization of this paper is as follows. In section \ref{sec:quantumapproach}, we construct the Lagrangian of axion electrodynamics in flat FRW universe and derive the resulting generalized TDHO Hamiltonian for the photons via canonical quantiztion.  In section \ref{sec:CB}, we obtain the phase of the photon quantum state  by means of the LR invariant method. We then analyze this phase and derive the polarization rotation from the geometric part of the phase. Finally, in section \ref{sec:analoguesim}, we present the analogue system.

\section{Canonical quantization of axion electrodynamics in flat FRW spacetime}\label{sec:quantumapproach}

The Lagrangian density for axion electrodynamics in curved spacetime is ($\hbar=c=1$)
\begin{equation}\label{eq:L}
  \mathcal{L}=-\sqrt{-g}\left(\frac{1}{4}F_{\mu\nu}F^{\mu\nu}+\frac{\beta}{2M}\phi F_{\mu\nu}\tilde{F}^{\mu\nu}\right)
\end{equation}
where $M$ (mass scale for broken global symmetry in Peccei-Quinn model \cite{Peccei1977}) has mass dimension 1, $\beta$ is a dimensionless coupling, $\phi$ is the pseudo-scalar axion field, $F^{\mu\nu}=\nabla^\mu A^\nu-\nabla^\nu A^\mu$ is the EM field tensor, and $\tilde{F}^{\mu\nu}=\frac{1}{2}\epsilon^{\mu\nu\rho\sigma}F_{\rho\sigma}$ is its dual. $\epsilon^{\mu\nu\rho\sigma}$ is the complete antisymmetric tensor, which is related to the absolutely antisymmetric Levi-Civita symbol $[\mu\nu\rho\sigma]$ through the metric according to: $\epsilon_{\mu\nu\rho\sigma}=\sqrt{-g}\, [\mu\nu\rho\sigma]$, $\epsilon^{\mu\nu\rho\sigma}=-(\sqrt{-g})^{-1}[\mu\nu\rho\sigma]$. For a spatially flat FRW universe the line element reads:
\begin{equation}
  ds^2=a^2(\eta)\left(-d\eta^2+d\bm{x}^2\right)
\end{equation}
where $a$ is the scale factor and $\eta$ is the conformal time. In the Coulomb gauge $A^\mu=(0,\bm{A})$ (the vector potential $\bm{A}$ satisfying $\nabla \cdot \bm{A}=0$), (\ref{eq:L}) becomes
\begin{equation}
  \mathcal{L}=\frac{1}{2}\,  \left(\bm{E}^2-\bm{B}^2\right)+\frac{2\beta}{M}\,\phi\,\bm{E}\cdot\bm{B}
\end{equation}
where $\bm{E}=-\partial_\eta \bm{A}$ and $\bm{B}=\nabla \times \bm{A}$ are the electric and magnetic fields, respectively. We write the Fourier transform of the vector potential as
\begin{equation}
   \bm{A}(\eta,\bm{x})=\int \frac{d^3 k}{(2\pi)^3}\sum_{\lambda}\bm{e}_{\bm{k}}^{(\lambda)}\chi_{\bm{k}}^{(\lambda)}(\eta)e^{i\bm{k}\cdot\bm{x}} \label{vecpot}
\end{equation}
where $\lambda=\pm$ runs over the two circular polarization states, $\bm{e}_{\bm{k}}^{(\lambda)}$ represent the corresponding polarization vectors, and $\bm{e}_{\bm{k}}^{(\lambda)\star}=\bm{e}_{-\bm{k}}^{(\lambda)},\, \chi_{\bm{k}}^{(\lambda)\star}=\chi_{-\bm{k}}^{(\lambda)}$. Substituting (\ref{vecpot}) in the Lagrangian $L=\int \mathcal{L} d^3 x$ and forming the row matrix $\bm{\chi}_{\bm{k}}^{(\lambda) T}=\left(\chi_{\bm{k}R}^{(\lambda)}\quad \chi_{\bm{k}I}^{(\lambda)}\right)$ from the real and imaginary parts of $\chi_{\bm{k}}^{(\lambda)}$,
 the Lagrangian for each Fourier mode and polarization state becomes ($L=\int \sum_\lambda \mathcal{L}_{\bm k}^{(\lambda)}(2\pi)^{-3}d^3 k$)
\begin{equation}
 \mathcal{L}_{\bm k}^{(\lambda)}=\frac{1}{2} \biggl(\dot{\bm{\chi}}_{\bm{k}}^{(\lambda) T}\dot{\bm{\chi}}_{\bm{k}}^{(\lambda)}-\omega_k^2\bm{\chi}_{\bm{k}}^{(\lambda)T}\bm{\chi}_{\bm{k}}^{(\lambda)}-\frac{4\beta \lambda \omega_k}{M}\phi \dot{\bm{\chi}}_{\bm{k}}^{(\lambda)T}\bm{\chi}_{\bm{k}}^{(\lambda)}\biggr)
\end{equation}
where dot denotes differentiation with respect to conformal time, and $\omega_k=|\bm{k}|$. The corresponding Hamiltonian is given by
\begin{equation}
\mathcal{H}_{\bm{k}}^{(\lambda)}=\bm{\pi}_{\bm{k}}^{(\lambda)T}\dot{\bm{\chi}}_{\bm{k}}^{(\lambda)}-\mathcal{L}_{\bm k}^{(\lambda)}, \ \ \ \ \ \
\bm{\pi}_{\bm{k}}^{(\lambda)T}=\partial \mathcal{L}_{\bm k}^{(\lambda)}/\partial \dot{\bm{\chi}}_{\bm{k}}^{(\lambda)}=\left(\pi_{\bm{k}R}^{(\lambda)}\quad \pi_{\bm{k}I}^{(\lambda)}\right)
\end{equation}
$\bm{\pi}_{\bm{k}}^{(\lambda)}$ being the conjugate momenta. Upon canonical quantization (operators represented by “hat”), we have
\begin{equation}\label{eq:generalizedhamitoni}
\hat {\mathscr{H}}_{\bm{k}}^{(\lambda)}=\frac{1}{2}\left [\hat {\bm{\pi}}_{\bm{k}}^{(\lambda)T}\hat{\bm{\pi}}_{\bm{k}}^{(\lambda)}+\omega_k^2  \left(1+\frac{4\beta^2}{M^2}\phi^2\right) \hat{\bm{\chi}}_{\bm k}^{(\lambda)T}\hat{\bm{\chi}}_{\bm k}^{(\lambda)}+\frac{2\beta \lambda \omega_k }{M}\phi \left(\hat{\bm{\pi}}_{\bm{k}}^{(\lambda)T}\hat{\bm{\chi}}_{\bm k}^{(\lambda)}+\hat{\bm{\chi}}_{\bm{k}}^{(\lambda)T}\hat{\bm{\pi}}_{\bm k}^{(\lambda)}\right)\right ]
\end{equation}
which is the Hamiltonian of a generalized TDHO of frequency $\omega_k$.

This Hamiltonian can be written as
\begin{equation}
  \hat{\mathscr{H}}_{\bm{k}}^{(\lambda)}=\left(\hat{\tilde{\bm{a}}}_{\bm{k}}^{(\lambda)T}\hat{\bm{a}}_{\bm{k}}^{(\lambda)}+1\right)\omega_k
\end{equation}
where the two-component column matrix
$$
  \hat{\bm{a}}_{\bm{k}}^{(\lambda)}(\eta)=\frac{1}{\sqrt{2\omega_k}}\left[\omega_k\hat{\bm{\chi}}_{\bm{k}}^{(\lambda)}+i\left(\hat{\bm{\pi}}_{\bm{k}}^{(\lambda)}+\frac{2\beta\lambda \omega_{k}}{M}\phi\hat{\bm{\chi}}_{\bm{k}}^{(\lambda)}\right)\right]
$$
and $\hat{\tilde{\bm{a}}}_{\bm{k}}^{(\lambda)T}=\left(\hat{a}_{\bm{k}1}^{(\lambda)\dag}\quad \hat{a}_{\bm{k}2}^{(\lambda)\dag}\right)$ are the standard particle annihilation and creation matrix operators, respectively. They  satisfy   ($i,j=1,2$)
\begin{eqnarray}\label{rel}
  \left[\hat{a}_{\bm{k}i}^{(\lambda)},\hat{a}_{\bm{k}j}^{(\lambda)\dag}\right]&=&\delta_{ij}  \nonumber \\
  (\hat{a}_{\bm{k}1}^{(\lambda)\dag}\hat{a}_{\bm{k}1}^{(\lambda)}+\hat{a}_{\bm{k}2}^{(\lambda)\dag}\hat{a}_{\bm{k}2}^{(\lambda)})\ket{n_{\bm{k}}^{(\lambda)},\eta}&=&{n_{\bm{k}}^{(\lambda)}}\ket{n_{\bm{k}}^{(\lambda)},\eta}
\end{eqnarray}
$\ket{n_{\bm{k}}^{(\lambda)},\eta}$ being the Hamiltonian eigenstate given by
\begin{equation}\label{Ham}
  \hat{\mathscr{H}}_{\bm{k}}^{(\lambda)}\ket{n_{\bm{k}}^{(\lambda)},\eta}=\left(n_{\bm{k}}^{(\lambda)}+1\right)\omega_k\ket{n_{\bm{k}}^{(\lambda)},\eta}.
\end{equation}
It follows that Hamiltonian (\ref{eq:generalizedhamitoni}) describes any number $n_{\bm{k}}^{(\lambda)}$ of photons of a given wavelength and polarization, interacting with the axion field. In the next section, we use the dynamical invariant method \cite{Lewis1969} to obtain the phase of their quantum state.

\section{Geometric Phase and Cosmological Birefringence}\label{sec:CB}

The invariant operator, which by definition is a constant of motion satisfying the von-Neumann equation, has been derived  for the generalized  TDHO Hamiltonian in \cite{Gao1990,Gao1991}. For Hamiltonian (\ref{eq:generalizedhamitoni}), it takes the form
\begin{equation}
  \hat{I}_{\bm{k}}^{(\lambda)}=\hat{\tilde{\bm{b}}}_{\bm{k}}^{(\lambda)T}\hat{\bm{b}}_{\bm{k}}^{(\lambda)}+1
\end{equation}
where
$$
  \hat{\bm{b}}_{\bm{k}}^{(\lambda)}(\eta)=\frac{1}{\sqrt{2}}\biggl\{\frac{1}{\rho_{k}^{(\lambda)}}\hat{\bm{\chi}}_{\bm{k}}^{(\lambda)}+i\biggl[\rho_{k}^{(\lambda)}\left(\hat{\bm{\pi}}_{\bm{k}}^{(\lambda)}+\frac{2\beta\lambda \omega_{k}}{M}\phi\hat{\bm{\chi}}_{\bm{k}}^{(\lambda)}\right)-\dot{\rho}_{k}^{(\lambda)}\hat{\bm{\chi}}_{\bm{k}}^{(\lambda)}\biggl]\biggr\}
$$
is the lowering and $\hat{\tilde{\bm{b}}}_{\bm{k}}^{(\lambda)T}=\left(\hat{b}_{\bm{k}1}^{(\lambda)\dag}\quad \hat{b}_{\bm{k}2}^{(\lambda)\dag}\right)$ is the raising matrix operator, and $\rho_{k}^{(\lambda)}(\eta)$ is an auxiliary variable that satisfies
\begin{equation}\label{eq:ro}
  \ddot{\rho}_{k}^{(\lambda)}+ \omega_{k}^2 \left(1-\frac{2\beta\lambda}{M\omega_{k}}\dot{\phi}\right)\rho_{k}^{(\lambda)}=(\rho_{k}^{(\lambda)})^{-3}.
\end{equation}
The components of the raising and lowering operators obey standard relations similar to (\ref{rel}) in terms of the eigenstates of the invariant operator, given by
\begin{equation}
  \hat{I}_{\bm{k}}^{(\lambda)}\ket{\bar{n}_{\bm{k}}^{(\lambda)},\eta}=\left(\bar{n}_{\bm{k}}^{(\lambda)}+1\right)\ket{\bar{n}_{\bm{k}}^{(\lambda)},\eta}.
\end{equation}
A photon state can be written as the superposition of the eigenstates of the invariant operator with phase factors (LR phase)  that follow from
\begin{equation}\label{eq:LRphase}
  \frac{d}{d\eta}\gamma_{\bar{n}_{\bm{k}}}^{(\lambda)}(\eta)=\bra{\bar{n}_{\bm{k}},\eta}i\partial_\eta-\hat{\mathscr{H}}_{\bm{k}}^{(\lambda)}\ket{\bar{n}_{\bm{k}},\eta}.
\end{equation}
This phase has been calculated for a generalized TDHO, with the result \cite{Gao1990}  
\begin{equation}\label{phase}
  \gamma_{\bar{n}_{\bm{k}}}^{(\lambda)}(\eta_0)=-\left(\bar{n}_{\bm{k}}^{(\lambda)}+1\right)\int_{0}^{\eta_0}d\eta \,(\rho_{k}^{(\lambda)})^{-2}.
\end{equation}
In our case, $\rho_{k}^{(\lambda)}$ obeys equation (\ref{eq:ro}), $\eta=0$ denotes the time when CMB photons left the last scattering surface and $\eta_0$ is the present time.

In the adiabatic limit of slow time variation, the auxiliary variable attains a definite form. In the auxiliary equation (\ref{eq:ro}), we introduce the adiabatic parameter $\epsilon$ ($\ll 1$) and write $\tau=\epsilon\eta$ to get
\begin{equation}\label{eq:adiaro}
  (\rho_{k}^{(\lambda)})^{-2}= \omega_k-\frac{\beta\lambda}{M}\epsilon\frac{d\phi}{d\tau}+O(\epsilon^2)\approx \omega_k-\frac{\beta\lambda}{M}\dot{\phi}.
\end{equation}
Adiabatic approximation, therefore, holds for $\beta\dot{\phi}/M\omega_k \ll 1$, i.e. for sufficiently slowly varying axion field. It follows also from differentiation of (\ref{eq:adiaro})  that in the adiabatic limit $\dot{\rho}_{k}^{(\lambda)}\approx 0$. Hence, $\hat{\mathscr{H}}_{\bm{k}}^{(\lambda)}\approx\omega_{\bm{k}}\hat{I}_{\bm{k}}^{(\lambda)}$, implying that the eigenstates of the invariant operator coincide with those of the Hamiltonian in this limit. Thus, $\bar{n}_{\bm{k}}^{(\lambda)}$  corresponds to the number of photons of a given wavelength and polarization, $n_{\bm{k}}^{(\lambda)}$. 

In adiabatic evolution, photons initially in an eigenstate  of the Hamiltonian will evolve into the instantaneous eigenstate after a time $\eta_0$, acquiring the phase $\gamma_{n_{\bm{k}}}^{(\lambda)}(\eta_0)$. Substitution of (\ref{eq:adiaro}) into (\ref{phase}) yields
\begin{equation}
  \gamma_{n_{\bm{k}}}^{(\lambda)}(\eta_0)=\left(n_{\bm{k}}^{(\lambda)}+1\right)
\left(\frac{\beta\lambda}{M}\Delta \phi-\omega_k\eta_0\right), \ \ \ \Delta\phi(\eta_0)=\phi(\eta_0)-\phi(0).
\end{equation}
The second term in above, which is immediately seen from (\ref{Ham}) to come from the second term in (\ref{eq:LRphase}), is the usual dynamical phase acquired along open curve $\mathcal{C}$ traversed by the state vector.
For any arbitrary quantum evolution, the geometric phase is the difference between the total phase and the dynamical phase \cite{Mukunda1993}. The total phase is the relative phase of the end point of $\mathcal{C}$ with respect to its starting point, as determined by the Pancharatnam criterion \cite{Pancharatnam1956}, and the dynamical phase is a locally additive functional of $\mathcal{C}$. Their difference is a gauge and reparametrization invariant phase that is only a functional of the image of $\mathcal{C}$ in the projective Hilbert space. 

Thus, for a general photon state that consists of $n_{\bm{k}}^{(+)}$ ($n_{\bm{k}}^{(-)}$) photons of positive (negative) helicity, the adiabatic noncyclic geometric phase acquired during its journey from the last scattering surface to the present epoch is given by
\begin{equation}\label{eq:adiageophase}
 \Gamma_{n_{\bm k}}^{\text{geom}}(\eta_0)=\frac{\beta}{M}\left( n_{\bm k}^{(+)}-n_{\bm k}^{(-)} \right)\Delta\phi(\eta_0).
\end{equation}
Therefore, a photon state with net helicity will exhibit the geometric phase shift (\ref{eq:adiageophase}) as it travels through the background slowly varying axion field. This is our main result. For a linearly polarized photon, being the superposition of two opposite circularly polarized photons, this will have the effect of a rotation of the polarization plane by $\beta \Delta\phi/M$, which corresponds to the classical result of \cite{Carroll1990,Carroll1991} for cosmological birefringence.

\section{The Analogue System}\label{sec:analoguesim}

Let us consider the possibility of viewing axion electrodynamics in curved spacetime as Maxwell electrodynamics in an effective medium in flat spacetime, the properties of which medium are determined by the metric of the original curved spacetime (see, e.g.\cite{Skrotskii1957,Mashhoon1973}).

The axion electrodynamics field equations obtained from Lagrangian (\ref{eq:L}) are
\begin{eqnarray}\label{eq:axionmaxwell}
\nonumber
   \nabla_\mu F^{\mu\nu}+\xi\nabla_\mu\left(\phi\tilde{F}^{\mu\nu}\right)&=&0\\
   \nabla_\mu \tilde{F}^{\mu\nu}&=&0
\end{eqnarray}
where $\xi=2\beta/M$. Consider a background flat spacetime with Cartesian coordinates, in which the electric and magnetic fields are defined using the decompositions $F_{\mu\nu}\rightarrow (\bm{E},\bm{B})$, and $\sqrt{-g}(F^{\mu\nu}+\xi\phi\tilde{F}^{\mu\nu})\rightarrow (-\bm{D},\bm{H})$. That is, define
\begin{eqnarray}
  E_i=F_{i0}, \quad D_i=\sqrt{-g}\left(F^{0i}+\xi\phi\tilde{F}^{0i}\right) \nonumber \\
  B_i=\frac{1}{2}\epsilon_{ijk}F_{jk}, \quad H_i=\frac{1}{2}\epsilon_{ijk}\sqrt{-g}\left(F^{jk}+\xi\phi\tilde{F}^{jk}\right). \label{constit}
\end{eqnarray}
In terms of these quantities, field equations (\ref{eq:axionmaxwell}) take the form
\begin{equation}
  \nabla \cdot\bm{D}=\nabla\cdot \bm{B}=0, \quad -\partial_0 \bm{D}+\nabla\times\bm{H}=\partial_0 \bm{B}+\nabla\times \bm{E}=0
\end{equation}
which are the source-free Maxwell equations in a material medium in flat spacetime. For flat FRW metric with cosmic time $dt=ad\eta$, the constitutive relations obtained from (\ref{constit}), thus, read:
\begin{eqnarray}
 \nonumber
  \bm{D} &=&a \bm{E}-\xi\phi \bm{B}, \\
  \bm{H} &=&a^{-1} \bm{B}+\xi\phi \bm{E}.
\end{eqnarray}
Rearranging these equations in the form of standard magnetoelectric equations, we have
\begin{eqnarray}\label{eq:constitutive}
\nonumber
  \bm{D} &=&\varepsilon \bm{E}+\alpha \bm{H} \\
  \bm{B} &=&\mu \bm{H}+\alpha \bm{E}
\end{eqnarray}
where $\varepsilon=a(1+\xi^2\phi^2)$ is the electric permittivity,  $\mu=a$ is the magnetic permeability, and $\alpha=-a\xi\phi$ is the magnetoelectric coefficient. Equations (\ref{eq:constitutive}) are the characteristic equations of bi-isotropic magnetoelectric media. Maxwell electrodynamics in this time-dependent medium presents the analogue system for Axion electrodynamics. 

 In 1948, Tellegen \cite{Tellegen1948} suggested that an assembly of randomly distributed parallel or antiparallel electric-magnetic dipole twins can furnish a new type of magnetoelectric material that is non-reciprocal and non-chiral. Since then, the very existence of the Tellegen media (now also known as axion media) in nature has been the subject of theoretical debate (see, e.g., \cite{Lakhtakia1994,Sihvola1995,Tretyakov2003}). However, there is experimental evidence that the natural composite chromium sesquioxide $\text{Cr}_{2}\text{O}_{3}$ can be modelled by a uniaxial Tellegen medium \cite{Hehl2008}. Moreover, the realization of an artificial isotropic Tellegen medium has been demonstrated by Gosh et al. \cite{Gosh2007,Gosh2008}.
In \cite{Gosh2008}, the authors have shown that by combining permanent electric and magnetic moments in particles, it is possible to construct a cross-correlation between the electric and magnetic properties of matter. These magnetoelectric particles interact with each other and their surroundings, and the nature of these interactions will determine the magnetoelectric response. By making this response decadently oscillatory (e.g. by applying an a.c. field with appropriate frequency) like the axion field  \cite{Kolb1990,Marsh2016}, one can construct the time-dependent medium defined by the constitutive equations (\ref{eq:constitutive}) and study EM wave propagation through it. This provides an analogue system for the simulation of cosmological birefringence.

\end{document}